\documentclass[twocolumn,superscriptaddress,showpacs,pre,aps]{revtex4-1}
\usepackage{graphicx}
\usepackage{natbib}
\usepackage{dcolumn}
\usepackage{bm}  % allows for index-generation
\begin{document}
%\mainmatter  % start of the contribution
\title{Voter model on a directed network: Role of bidirectional opinion exchanges}

\author{Sung-Guk Han}
\affiliation{BK21 Physics Research Division and Department of Physics,\\
Sungkyunkwan University, Suwon 440-746, Korea}

\author{Jaegon Um}
\affiliation{School of Physics, Korea Institute for Advanced Study, Seoul 130-722, Korea}

\author{Beom Jun Kim}
\email[Corresponding author:]{beomjun@skku.edu}
\affiliation{BK21 Physics Research Division and Department of Physics,\\
Sungkyunkwan University, Suwon 440-746, Korea}

%\date{\today}
\begin{abstract}        % give a summary of your paper
The  voter model with the node update rule is numerically investigated on a
directed network.  We start from a directed hierarchical tree, and split and
rewire each incoming arc at the probability $p$.  In order to
discriminate the better and worse opinions, we break the $Z_2$  symmetry
($\sigma = \pm 1$)  by giving a little more preference to the opinion $\sigma =
1$. It is found that as $p$ becomes larger, introducing more complicated
pattern of information flow channels, and as the network size $N$ becomes
larger, the system eventually evolves to the state in which more voters agree on
the better opinion, even though the voter at the top of the hierarchy
keeps the worse opinion. We also find that the pure hierarchical tree
makes opinion agreement very fast, while the final absorbing state can easily
be influenced by voters at the higher ranks. On the other hand, although the
ordering occurs much slower, the existence of complicated pattern of
bidirectional information flow allows the system to agree on the better opinion.
\end{abstract}
%\keywords {opinion dynamics, voter model, directed network, hierarchy}
% 89.75.-k : Complex systems
% 87.23.Ge : Dynamics of social systems
% 89.75.Fb : Structure and organization in complex systems
%\keywords{Suggested keywords}%Use showkeys class option if keyword
                              %display desired

\maketitle              % typeset the title of the contribution

Recently, social phenomenon of the opinion formation has been studied
through the uses of simple prototypical models such as 
the Axelord's culture dissemination model~\cite{axelrod} and the 
voter model~\cite{41,50,42-51-52-53,54,55,59,43,45,57}.
In particular, the voter model on complex networks 
has been popularly studied~\cite{41,50,42-51-52-53,54,55,59}, reflecting
the broad research interest in dynamic behaviors of complex networks~\cite{doro}.
In the language of statistical physics, the voter model has the same
symmetry as the Ising model if the Ising spin is replaced by the 
opinion of a voter who can have two different opinions $\sigma = \pm 1$. 
Likewise, the opinion formation dynamics in the voter model corresponds to 
the coarsening process in  the Ising system, although the former is driven by
the interface noise while the latter by the surface tension in $d$-dimensional
lattice~\cite{57}, resulting in interesting differences~\cite{57,59}. The
analytic approaches have been successfully
applied for the voter model in the regular $d$-dimensional lattice~\cite{43}
and in the complex networks~\cite{54,50,41}: in $d$-dimensional regular lattice, 
the density $\rho$ of active bonds connecting opposite opinions decays to zero only
for $d \leq 2$, whereas for $d > 2$ the system never approaches completely ordered 
absorbing state~\cite{43,57}. The voter model on fractal structures with $d < 2$
also exhibits the complete ordering~\cite{45}. 
The ordering dynamics of the voter model in various complex networks share
common features~\cite{59,54,42-51-52-53,50,41}: if started from randomly
distributed opinions, the system first approaches a quasi-stationary state and
stays there for a life time $\tau$.  As time is elapsed further the system
approaches the absorbing state characterized by $\rho = 0$.
Interestingly, the life time $\tau$ of the disordered quasi-stationary state
increases with the network size $N$~\cite{59,54,42-51-52-53,50,41}, which makes
the complete ordering in the thermodynamic limit impossible, in accord with the
study for regular lattices in a very high dimensionality~\cite{43}.
There is now growing consensus that the ordering dynamics of the voter model
depends only weakly on the topological details of the underlying complex 
networks~\cite{59,50}, and that the first and the second moments of the
degree distribution mostly determine the ordering behaviors for networks
with negligible degree-degree correlation~\cite{54,50}.
We emphasize that most existing studies of the voter
and other related models have assumed that the networks are undirected.
The most recent studies 
for the ordering dynamics of the voter model have started to consider 
the directedness of networks~\cite{Masuda,Serrano}. 

\begin{figure}
\includegraphics[width=0.42\textwidth]{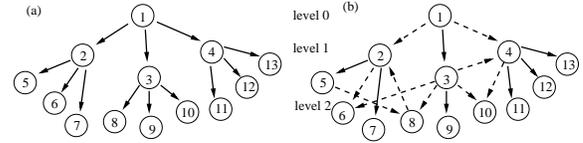}
\caption{\label{fig:tree}
(a) A directed tree network with the perfect hierarchical structure  
is initially built with the branching ratio 3 and the number of levels
$L=3$. The total number of vertices is given by $N = (3^{L}-1)/2$.  
(b) Each incoming edge is visited one by one, and then split 
into two edges at the probability $p$: the existing one but with the half weight and the 
new one of the half weight from randomly chosen other vertex.
Solid (dashed) lines denote the incoming edges of the weight unity
(half).
}
\end{figure}

\begin{figure*}
\includegraphics[width=0.85\textwidth]{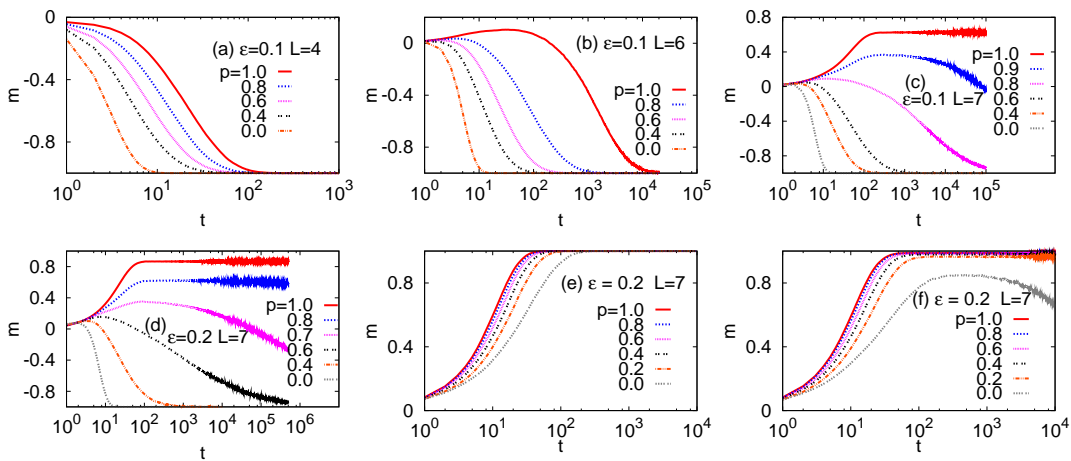}
\caption{\label{fig:m}
The order parameter $m(t)\equiv (1/N)\langle\sum_v\sigma_v\rangle$ averaged
over 500-2000 independent runs versus the Monte-Carlo time step $t$ for 
directed network of sizes (a) $L=4$ ($N=40$), (b) $L=6$ ($N=364$), (c) $L=7$ ($N=1093$) 
with $\epsilon=0.1$ and (d) $L=7$ with $\epsilon=0.2$ 
at various values of the splitting probability $p$. For
small sizes, the voters favor the opinion of the top $\sigma = -1$ 
at any value of $p$. However, at sufficiently large values of $p$ and for
larger sizes, finite fraction of voters begin to have
opinions $\sigma=1$ against the top vertex, for a very long time.
(e) Undirected network of size $L=7$ ($N=1093$) with $\epsilon=0.2$ in which all edges are undirected. 
(f) Different version of undirected
network of size $L=7$ ($N=1093$) with $\epsilon=0.2$ in which all edges are undirected except the ones 
that connect the top voter and her lower rank direct acquaintances.
}
\end{figure*}

In the sociological and biological disciplines, directed networks are
abundant. To name a few, the world wide web, the email communication 
network, and transcriptional regulatory networks are described better
as directed networks. In such directed networks, one generally expects that 
dynamical behaviors such as synchronization of phase oscillators,
opinion formation in the voter model, and the epidemic spread of diseases,
exhibit sharp differences from those in undirected networks. 
In Ref.~\onlinecite{55}, both the Hodgkin-Huxley model in neuroscience
and the sociological voter model have been studied in the small-world
type directed network: as more edges become directed,
there occur clear changes in dynamic behaviors, and it has been found
that bidirectional information exchange helps consensus state
to emerge in the voter model.
S{\'a}nchez {\it et. al.}~\cite{49} found that 
the directed links on the directed 
small-world networks play an important role altering the nature of the phase
transition. These findings show that the directedness of networks strongly 
affects dynamical and critical behaviors of the system. 

In this paper we study a sociological game model of voters connected through
the directed network structure (see Fig.~\ref{fig:tree}).  
Our directed network
originates from the perfect hierarchical tree structure, which has been found
to yield the best synchronizability for the set of identical
oscillators~\cite{nishikawa} but not for the nonidentical oscillators~\cite{jaegon}.  
In order to build directed networks, we start
from the tree with $L$ layers and branching ratio $b$ [$L=3$, $b=3$ in
Fig.~\ref{fig:tree}(a)].  We then sequentially visit each incoming edge of the
weight unity and with the probability $p$ reduce its weight to 1/2 and make a
new incoming edge of the weight 1/2 from a randomly chosen vertex. It is to be
noted that our way of splitting edges conserves the incoming strength for each
vertex and the resulting directed network contains the original tree.  When the
above procedure is repeated for all incoming edges in the original tree, the
resulting network can possess  loops [e.g., $2 \rightarrow 5\rightarrow
8\rightarrow 2$ in Fig.~\ref{fig:tree}(b)] and directed upward edges [e.g.,  $8
\rightarrow 2$ in Fig.~\ref{fig:tree}(b)]. 
Consequently, our network model embeds the hierarchical directed network, which
exists abundantly in human societies: Decision makings in corporations,
governments, and religious organizations can be described as opinion formation on
hierarchical structures~\cite{Arbesman}.  

After the construction of directed networks, we perform
the voter model simulations on top of the network structures. The original
voter model with two opinions $\sigma = \pm 1$ for undirected 
networks proceeds as follows:  
A vertex $v$ is chosen at random.  One of its neighbors, say $w$, 
is chosen randomly among $v$'s direct acquaintances. The opinion
$\sigma_v$ of the vertex $v$ is changed to that of $w$s. The $N$
repetitions of the above process constitutes a unit time step so that
each voter has chance to update its opinion once on average.  
The generalization for directed and weighted networks is straightforward:
we only need to change the procedure so that $w$ is now chosen
only from $v$'s incoming neighbors ($w \rightarrow v$) and 
the weight of this chosen incoming link is used as the probability
of acquaintance of $v$ and $w$.
Suppose that we play such a voter model on  the directed tree structure 
in Fig.~\ref{fig:tree}(a). 
The final absorbing state of the whole system is solely determined by
the opinion of the
top vertex since it does not have incoming edge and thus no one can
persuade him otherwise~\cite{fnote1}. 
In other words, if the
top vertex has made somehow an incorrect decision, the whole society of
voters cannot escape from the disastrous situation in which all voters have
the same incorrect opinion of the top vertex. However, in a real experience,
one observes that it is plausible that a large scale of consensus of
opinions can be made although it is against the opinion of the top.
In this regard, we give a distortion in the original voter model as
follows: (i) the top voter is assigned the opinion $\sigma_1 = -1$ which
we assume as an incorrect opinion (or poorly decided opinion). All other
voters are assigned either 1 or $-1$ randomly at the initial stage. 
(ii) A voter $v$ is chosen at random, and one of 
its incoming neighbor $w$ is also randomly picked. At the probability
of the weight (either 1 or 1/2 in our model; see Fig.~\ref{fig:tree}) 
for the link connecting the two, the following procedure is performed. 
(iii) If $\sigma_w(t) = -1$ at time $t$,  $v$ changes its opinion to 
$w$'s one, i.e., $\sigma_v(t+1)  = -1$  
at the probability $1-\epsilon$. On the other hand, if $\sigma_w(t) = 1$,
$v$ always follows $w$'s opinion, i.e., $\sigma_v(t+1) = 1$ at the probability
unity.  In words, the better opinion $\sigma = 1$ is always accepted
while the slightly worse opinion $\sigma = -1$ is accepted with a bit
of hesitation.
Within this modified voter model, the original 
$Z_2$ symmetry is broken (compare with Ref.~\onlinecite{57}) 
by the asymmetry parameter $\epsilon$ 
and we drive the system to be in favor of the opinion $\sigma=1$, by giving each voter a 
slight preference toward $\sigma =1$. The incoming strength, i.e., the
sum of the weights of incoming arcs, is conserved in our model, which we
interpret as that the total time or resources each voter can use to make
contact is fixed to a constant. We have also used the unweighted version 
of our modified voter model in which 
the above step (iii) is performed always for the pair $(v,w)$, regardless
of the weight of the link. We have observed only insignificant differences
and all the results presented in this Brief Report are for the voter model with
the edge weight taken into account.

The motivation of our voter model is to mimic real-world situations in 
which there are two competing options or choices and one is intrinsically
better than the other. The worse option can still pervade the
whole system, if a dominant agent strongly advocates it. In our model, 
the intrinsically better option is described by the symmetry breaking
parameter $\epsilon > 0$, and the strong advocate of the worse option
is represented as $\sigma = -1$ assigned for the top voter.
The purpose of our Brief report is then to reveal what can be the effects
of the connection structure to change the situation so that the better
option $\sigma = 1$ manages to spread across the whole society.
\begin{figure}
\includegraphics[width=0.38\textwidth]{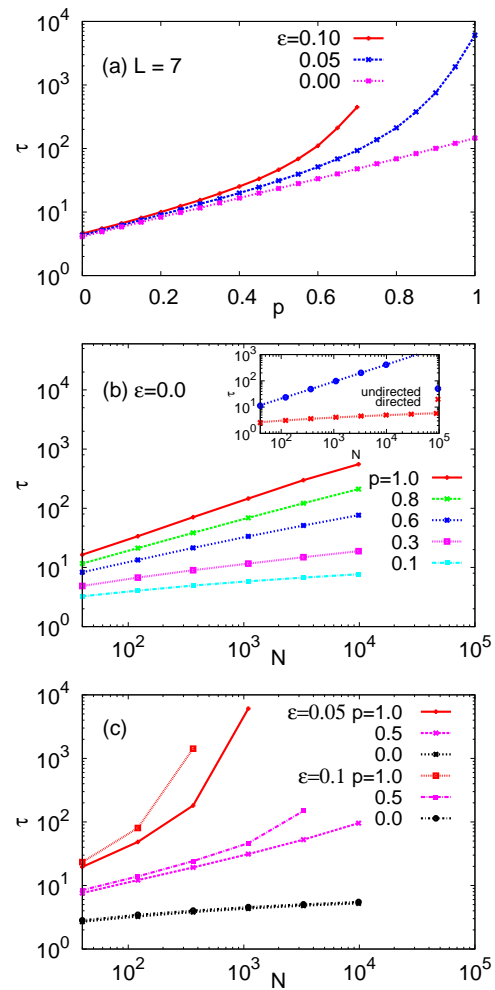}
\caption{\label{fig:time}
Relaxation time $\tau$ versus (a) $p$ and [(b) and (c)] $N$. 
(a) $\tau$ monotonically increases with $p$. 
(b) For $\epsilon=0.0$, at higher values of $p$, $\tau$ grows with $N$ 
algebraically, whereas at lower values of $p$ $\tau$ grows with $p$
logarithmically. The inset exhibits the result of $\tau$ vs. $N$ 
for directed tree and undirected tree (i.e., where $p=0$) and 
lines are results from the least-square-fit to algebraic and logarithmic
functions, respectively.
(c) For $\epsilon=0.05$ and $0.1$ at higher values of $p$,
$\tau$ increases with $N$ faster than algebraic.
}
\end{figure}

We construct directed networks of various sizes as described above and perform
the simulation of our modified voter model. During the simulation, we measure
the order parameter defined as $m(t) \equiv (1/N)\langle \sum_v \sigma_v(t)
\rangle$ with $\langle \cdots \rangle$ being the average over network
realizations and initial configurations. The consensus state with completely
agreed opinion is characterized by either $m = 1$ or $=-1$.  When the splitting
probability $p$ in Fig.~\ref{fig:tree} is sufficiently small, we expect that
the system approaches the consensus state in which all voters have opinion $\sigma=-1$
since the unidirectional information flow is abundant, and very few voters can
disobey the opinion of the top voter. In contrast, as $p$ is increased further
and thus as lots of directed loops are newly formed, one expects that the
situation reverses and a large number of voters will favor the
opinion $\sigma=1$. In Fig.~\ref{fig:m}, 
the order parameter is displayed as a function of $t$ 
for various sizes (a) $L=4$ ($N=40$), (b) $L=6$ ($N=364$), and (c) $L=7$ ($N=1093$) 
at $\epsilon =0.1$, and for comparison (d) $L=7 (N=1093)$ at $\epsilon=0.2$ in directed
networks, and the undirected cases of (e) and (f) of networks of size $L=7$
 ($N=1093$) at $\epsilon=0.2$ are discussed later.
It is clearly seen that the above expectation becomes indeed correct as the
system size is increased. For small sizes, the system eventually approaches to
the absorbing state $m=-1$ at any value of the splitting probability $p$. However,
as the size becomes larger [see Fig.~\ref{fig:m}  
(c) and (d) for
$L=7$ ($N=1093$)] $m$ increases toward positive values at higher values of $p$  
, indicating that there
are more voters with $\sigma = 1$ than in the initial state, and $m$ stays at an 
almost constant value for a very long time~\cite{fnote2}.
In the later stage, $m$ appears
to decay to $-1$ eventually, very slowly. 
Furthermore, the larger value of $\epsilon$ makes more voters agree on $\sigma = 1$
at lower values of $p$ as shown in Figs.~\ref{fig:m} (c) and ~\ref{fig:m} (d). 
The comparison of the time decay behaviors for different sizes makes us conclude that
for even bigger sizes, the system will stay at positive values of $m(t)$ in
an extremely long time. Due to the limitation of computing resources, it is 
difficult to decide whether $m(t\rightarrow \infty)=1$ or $-1$ for large
enough value of $p$.
%Relaxation time $\tau$ versus (a) $p$ and (b) $N$.
%(a) The relaxation time $\tau$ monotonically increases with $p$. 
%(b) For higher values of $p$, $\tau$ grows with $N$ exponentially.
%$\epsilon =0.1$ has been used. 

Another interesting observation one can make from Fig.~\ref{fig:m}
(c) and (d) is that the approach toward $m = -1$ for $p=0$ occurs much faster 
than  toward a positive value of $m$ for $p=1$.
This indicates that although the system can achieve the state in which
majority of voters keep the better opinion, it takes longer time to get there.
We next compute the active bond density $\rho(t) = [\sum_{i}\sum_{j \in {\cal V}^{in}_{i}}
(1-\sigma_{i}\sigma_{j})/2]/\sum_{i}|{\cal E}^{in}_i|$, where
${\cal V}^{in}_{i}$ (${\cal E}^{in}_i$) is the set of the incoming nearest vertices (edges) 
of the $i$th voter and $|{\cal E}^{in}_i|$ is the number of edges in the set ${\cal E}^{in}_i$.
The relaxation time to reach the absorbing state $\rho(t)=0$ or $m(t)=-1$ is also calculated.
We observe that as $p$ is increased, $\tau$ also increases monotonically, which 
indicates that the complete opinion agreement to $\sigma = -1$ takes longer time as the network 
structure becomes more different from the hierarchical structure, and
$\tau$ increases more rapidly for the higher value of $\epsilon$,
as shown in Fig.~\ref{fig:time} (a). 
For $\epsilon=0.0$, $\tau$ increases in a logarithmic way with $N$ 
at $p=0.0$, whereas in undirected tree the relaxation time increases 
algebraically with $N$ as displayed in the inset of Fig.~\ref{fig:time} (b), 
implying that the consensus of opinions occurs much faster in directed tree networks
than in undirected ones. It is notable that the crossover from the logarithmic behavior to
the algebraic one exists as $p$ is increased in the case of
$\epsilon=0.0$ [see Fig.~\ref{fig:time} (b)]. 
Consequently, we conclude that the bidirectional information flows make the agreement of 
opinion slower through the formation of a large number of voters against
the opinion of the top voter. Figure~\ref{fig:time} (c) shows that when $\epsilon >0$,     
$\tau$ increases very fast with $N$ for higher values of $p$ (faster than
the algebraic increase).

%{\bf removed: Figure 4 shows how $m$ changes in time for the higher value
%of the asymmetry parameter ($\epsilon = 0.2$) for the system sizes 
%$L=6$ and $L=7$ [compare with Fig.~\ref{fig:m}(c) and (d)].
%From the comparison with Fig.~\ref{fig:m} for $\epsilon = 0.1$,
%the larger value of $\epsilon$ makes more voters agree on $\sigma = 1$
%at lower values of $p$.  We conclude that $m$ decays 
%toward $m=-1$ in a {\it logarithmic} way 
%and that the relaxation time
%toward the absorbing state $m=-1$  increases {\it exponentially} with $N$.
%In contrast, the voter model in various complex networks is known to
%have an {\it exponential} decay to absorbing state, after the
%relaxation time $\tau$ which increases {\it algebraically} with $N$.} 

For comparisons, we also study the undirected network version 
of our model in which the network is constructed in the same way as for directed case (see
above) and then all edges are assumed to be symmetric, i.e., bidirectional. We
observe that the final consensus state $m(t \rightarrow \infty) > 0$ is reached
as far as $\epsilon >0$, regardless of the splitting probability $p$, which leads 
us to conclude that our voter model in directed network and undirected network
exhibit a quite different behavior [see Figs.~\ref{fig:m} (e) and ~\ref{fig:m} (f)]. 
Nevertheless, our main conclusion of the role
of the bidirectional information flow is still valid in the undirected
networks: the more abundant bidirectional opinion exchanges
help the majority of voters to agree on the better opinion.

In summary, we have investigated the simple sociological model of voters
on directed networks. Starting from a directed tree structure,
each edge weight is halved and a new incoming edge with a half weight is newly
connected at the probability $p$.  As more edges are split, making the network
possess more complicated pattern of information flow channels, it has been
found that finite fraction of voters can hold opinions against the opinion of
the root voter at the top. This trend becomes more evident as the asymmetry
between the two opinions $\sigma = \pm 1$  becomes bigger, making one opinion
more favorable than the opposite, and also as $p$ and $N$ become larger. The
approach toward the absorbing state has been found to be very slow for large
values of $p$. It has also been found that the relaxation time increases very 
fast with the size $N$.  
%Our results need to be compared with the corresponding results
%obtained for the voter model on undirected complex
%networks~\cite{41,50,54,42-51-52-53,59}, where it has been found that the
%relaxation time increases algebraically with $N$. 

We acknowledge the support from the Korea Science and Engineering 
Foundation (Grant No. R01-2007-000-20084-0) and the support from Asia 
Pacific Center for Theoretical Physics (APCTP).
% ---- Bibliography ----


\begin{thebibliography}{6}
\bibitem{axelrod} R. Axelrod, J. Conflict Resolut. {\bf 41}, 203 (1997);
C. Castellano, M. Marsili, and A. Vespignani, Phys. Rev. Lett. {\bf 85},
3536 (2000); K. Klemm, V. M.  Egu{\'i}luz, R. Toral, and M. San Miguel,
Phys. Rev. E {\bf 67}, 026120 (2003).
\bibitem{41} D. Vilone and C. Castellano, Phys. Rev. E {\bf 69}, 016109
(2004).
\bibitem{50} F. Vazquez and V. M.  Egu{\'i}luz, New J. Phys. {\bf 10},
063011 (2008).
\bibitem{54} V. Sood and S. Redner, Phys. Rev. Lett. {\bf 94}, 178701
(2005).
\bibitem{42-51-52-53} C. Castellano, D. Vilone, and A. Vespignani,
Europhys. Lett. {\bf 63}, 153 (2003); X. Castell{\'o}, R. Toivonen, V. M.
Egu{\'i}luz, J.
Saram{\"a}ki, K. Kaski, and M. San Miguel, {\it ibid.} {\bf 79}, 66006
(2007);  K.
Suchecki, V. M. Egu{\'i}luz, and M. San Miguel, {\it ibid.}  {\bf 69},
228
(2005);  K. Suchecki, V. M. Egu{\'i}luz, and M. San Miguel, Phys. Rev. E
{\bf
72}, 036132 (2005).
\bibitem{55} S. M. Park and B. J. Kim, Phys. Rev. E {\bf 74}, 026114
(2006).
\bibitem{59} C. Castellano, V. Loreto, A. Barrat, F. Cecconi, and D.
Parisi,
Phys. Rev.  E {\bf 71}, 066107 (2005).
\bibitem{57} I. Dornic, H. Chat{\'e}, J. Chave, and H. Hinrichsen,
Phys. Rev. Lett. {\bf 87}, 045701 (2001).
\bibitem{43} L. Frachebourg and P. L. Krapivsky, Phys. Rev. E {\bf 53},
R3009 (1996).
\bibitem{45} K. Suchecki and J. A. Holyst, Physica A {\bf 362}, 338
(2006).
\bibitem{doro} See, e.g., S. N. Dorogovtsev, A. V. Goltsev, and J. F. F.
Mendes, Rev. Mod. Phys.
{\bf 80},  1275 (2008) and references therein.
%\bibitem{58} D. Boyer and O. Miramontes, Phys. Rev. E {\bf 67},
%035102(R)(2003).
\bibitem{Masuda} N. Masuda and H. Ohtsuki, New J. Phys. {\bf 11}, 033012
(2009).
\bibitem{Serrano} M. {\'A}. Serrano, K. Klemm, F. Vazquez, V. M.
Egu{\'i}luz, and M. San Miguel, J. Stat. Mech.: Theory Exp. (2009),
P10024.
\bibitem{49} A. D. S\'anchez, J. M. L\'opez, and M. A. Rodr\'iguez,
Phys.
Rev. Lett. {\bf 88}, 048701 (2002).
\bibitem{nishikawa} T. Nishikawa and A. E. Motter, Phys. Rev. E {\bf
73},
065106(R) (2006).
\bibitem{jaegon} J. Um, S.-I. Lee, and B. J. Kim,  J. Korean Phys. Soc.
{\bf 53}, 491 (2008).
\bibitem{Arbesman} See S. Arbesman, J. M. Kleinberg, and S. H. Strogatz,
Phys. Rev.  E {\bf 79}, 016115 (2009), and references therein.
%\bibitem{Mobilia} M. Mobilia, Phys. Rev. Lett. {\bf 91}, 028701 (2003).
\bibitem{fnote1}Similarly, effects of the existence of a 
single zealot who has more preference for a particular opinion
have been studied in M. Mobilia, Phys. Rev. Lett. {\bf 91}, 028701 (2003).
\bibitem{fnote2}{We also measure the 
order parameter $m_S(t)$ defined as the average  over surviving runs,
i.e., the time series that decays to the absorbing state $\sigma = -1$ 
are excluded. As $p$ is increased, $m_S(t)$ favors a positive value similarly
to $m(t)$. Interestingly, in a broad range of $p$, $m_S(t)$ first decreases and 
then increases, before it eventually decays to $-1$ .}
\end{thebibliography}
\end{document}